\documentclass[a4paper,11pt,oneside]{article}
\usepackage{graphicx}
\usepackage{amsmath}

\newcommand{\ds}[1]{\displaystyle{#1}}

\newcommand{\Z}{\mathbf Z}

\newcommand{\bq}{\begin{quote}}
\newcommand{\eq}{\end{quote}}

\newcommand{\ub}{\underbrace}
\newcommand{\cl}{\centerline}

\newcommand{\vs}{\vskip0.3cm\noindent}

\newcommand{\dequ}{\begin{equation} \vspace{3mm}}
\newcommand{\fequ}{\vspace{3mm} \end{equation}}

\newcommand{\ba}{\begin{array}}
\newcommand{\ea}{\end{array}}
\newcommand{\bc}{\begin{center}}
\newcommand{\ec}{\end{center}}
\newcommand{\bi}{\begin{itemize}}
\newcommand{\ei}{\end{itemize}}
\newcommand{\ben}{\begin{enumerate}}
\newcommand{\een}{\end{enumerate}}

\newcommand{\bfi}{\begin{figure}[hbtp]}

\newcommand{\efi}{\end{figure}}

\newcommand{\dr}{\partial}
\newcommand{\beq}{\begin{equation}}
\newcommand{\eeq}{\end{equation}}
\newcommand{\beqar}{\begin{eqnarray}}
\newcommand{\eeqar}{\end{eqnarray}}

\renewcommand{\bar}{\overline}
\newcommand{\bit}{\begin{itemize}}
\newcommand{\eit}{\end{itemize}}

\newcommand{\w}{\omega}

\newcommand{\eps}{\varepsilon}

\renewcommand{\vs}{\vskip0.1cm\noindent}

\renewcommand{\bq}{\begin{quote}}
\renewcommand{\eq}{\end{quote}}

\newcommand{\tensna}{\bar{\bar{\nabla}}}

\newcommand{\Ta}{\mbox{Ta}}

\newcommand{\St}{\mbox{St}}
\newcommand{\Rep}{\mbox{Re}}

\renewcommand{\ds}{\displaystyle}

\newcommand{\CC}{\mbox{\small (C)}}
\renewcommand{\a}{\mbox{a}}
\renewcommand{\S}{${\cal S}\,\,$}

\begin{document}

\cl{\bf Asymptotic analysis of chaotic particle sedimentation}
\cl{\bf and trapping 
in the vicinity of a vertical upward streamline.}
\vs\vs
\cl{\bf Jean-R\'egis Angilella\footnote{Present Address : LAEGO, rue du Doyen Marcel Roubault BP 40
   54501 Vand\oe uvre-l\`es-Nancy, France.}}

\cl{Nancy-Universit\'es, LEMTA, CNRS UMR 7563}
\cl{2 avenue de la For\^et de Haye, 54504 Vand\oe uvre-les-Nancy, France.}

\vskip.5cm 

\cl{\bf Abstract}
\vskip.1cm
 The sedimentation of a heavy Stokes particle in a laminar plane or axisymmetric  flow 
 is investigated by means of asymptotic methods. We focus on the occurrence of 
Stommel's retention zones, and on the splitting of their separatrices.
The goal of this paper is to analyze under which conditions these retention zones can
form, and under which conditions they can break and induce chaotic particle
settling.
The terminal velocity of the particle in still fluid is of the order of
the typical velocity of the flow, and the particle response time is much smaller 
than  the typical flow time-scale. It is observed that if  the  flow is steady and 
has an upward streamline where the vertical velocity has a strict 
 local maximum, then inertialess particle trajectories can take locally the form of elliptic 
Stommel cells, provided the particle terminal velocity is close enough to the local peak flow velocity. 
These structures  only depend on the local
flow topology  and do not require the flow to have closed streamlines 
or stagnation points.
If, in  addition, the flow is submitted to a weak 
time-periodic perturbation,
classical expansions enable one to
write the particle dynamics as a hamiltonian system 
with one degree of freedom, plus a perturbation containing both
the dissipative terms of the particle motion equation and the flow
unsteadiness.
Melnikov's
method therefore provides accurate criteria to predict the splitting
of the separatrices of the elliptic cell mentioned above, 
leading to chaotic particle trapping
and chaotic settling. The effect of particle inertia and flow unsteadiness
on the occurrence of simple zeros in Melnikov's function is discussed.
Particle motion in a  plane cellular
flow and in a vertical pipe is then investigated to illustrate these results.

\vskip.5cm

\section{Introduction}

The motion of tiny heavy particles in  either laminar 
or turbulent flows is of major importance for 
the understanding of many industrial or natural flows.
Even when particles do not significantly modify the surrounding flow,
this topic shows many unsolved questions. 
When the flow is turbulent,  modeling   the dispersion and sedimentation of
heavy particles is a difficult task which has retained much attention since
the pioneering works by Csanady \cite{Csanady1963},
 Snyder \& Lumley \cite{Snyder1971} or Maxey \cite{Maxey1987}, to name but a few.
In laminar flows, particle motion is also investigated for at least two reasons.
First, a better understanding of particle dispersion and sedimentation in fully
developped turbulence can be achieved if one understands the detailed interactions
between particles and elementary flow structures like vortices or shear layers.
Secondly, even in laminar flows, the behaviour of
small heavy particles can be rather unexpected,
and chaotic settling or permanent suspension (``trapping'') can occur.
In the limit of vanishing particle Reynolds number,
these complex trajectories are not due to 
wake effects such as vortex shedding, since 
the disturbance flow set up
 by the inclusion obeys the linear Stokes equation. They are
due to the gradients of the unperturbed fluid flow, which introduce severe
non-linearities into the particle motion equations.

\vs

Experiments and calculations conducted since 1900 revealed 
that passive heavy Stokes particles dropped in a cellular flow 
can be trapped in the vicinity of vertical upward streamlines 
and   remain suspended for a long time, under the combined effect
of gravity and of the hydrodynamic force (B\'enard \cite{Benard1900},
Stommel \cite{Stommel1949},
Simon \& Pomeau \cite{Simon1991},  Cerisier {\it et al.}
\cite{Cerisier2005}, Tuval {\it et al.} \cite{Tuval2005}).
Such nearly closed trajectories are often called  ``Stommel retention zones'',
even though they are rather loosely defined. They are the central point of the present work.
Stommel investigated these structures to understand plankton
settling in a lake, where wind-induced cellular motion is
likely to induce strong inhomogeneities in plankton concentration.
In the last decades, various numerical analyses
 revealed that aerosol sedimentation
can display  complex features, whether chaotic or not,  in elementary 
{ laminar} flows such as plane cellular flow (Maxey \& Corrsin
\cite{Maxey1986},
  Fung \cite{Fung1997}, Rubin {\it et al.} \cite{Rubin1995}), or ABC flow 
(Mac Laughlin 
\cite{McLaughlin1988}).
Some of these analyses revealed that the occurrence of  Stommel cells
in various parts of the flow
played a key role in  complex particle motions. For example,
some particles released in a plane cellular flow, submitted to a weak time-periodic
perturbation, were observed to  wander around some kind of Stommel cells,
and could also be trapped temporarily within these cells,
in a rather chaotic manner \cite{Fung1997}. The occurrence of Stommel cells, as well as
their role in chaotic sedimentation, is therefore a topic of interest to understand and
quantify complex particle motion in simple flows.

The goal of the present paper is to derive
 analytical criteria giving the particle parameters for which
chaotic sedimentation and/or trapping occurs,
 for  heavy isolated Stokes particles dropped in the vicinity of a
vertical streamline where Stommel retention zones are likely to form.
Two-dimensional and 
axisymmetric
divergence-free
flows will be considered.  
The  paper is organized as follows.
First, particle dynamics
in the vicinity of an upward streamline of any steady 2D flow is investigated
(section \ref{H}). Conditions leading to the formation of retention zones
are derived.
Then the splitting of the
separatrices of these retention zones is analyzed by means of Melnikov's method 
(section \ref{melnik}).
This analysis is applied to chaotic  sedimentation in a plane cellular 
flow in section \ref{appli}.
 Finally, these results are generalized to
axisymmetric flows ( section \ref{3D}). In the next few paragraphs
we present the basic equations of particle dynamics used in this paper.

\vs

 If $\vec X_p(t)$ is the particle position 
at time $t$,
the simplest motion equation for heavy isolated Stokes particles in
a fluid of infinite extent is
$$
m_p {\vec{\ddot X}_p} = m_p \vec g + 6\pi\mu_f\,{\a} \Big( \vec V_f(\vec X_p,t) -  \vec{\dot X}_p  \Big),
$$
where $m_p$ and a denote the particle mass and radius respectively, 
$\mu_f$ is the fluid dynamical viscosity, and $\vec V_f$ is the fluid velocity field
in the absence of particle.
 This equation is valid provided the particle Reynolds,
Stokes  and Taylor numbers  are much smaller than unity :
$$
\Rep =  \frac{\a |\vec{\dot X}_p-\vec V_f|}{\nu}\ll 1,\quad \mbox{and}\quad 
  \St = \frac{\a^2 \w_1}{\nu}\ll 1,
$$
$$
\mbox{and}\quad 
  \Ta = \frac{\a^2 |\nabla \vec V_f|}{\nu} \ll 1,
$$
 so that the disturbance flow
 due to the inclusion is  a creeping  quasi-steady
 flow 
($\w_1$ is the  typical time-scale of this flow).
In addition, the particle density $\rho_p$
  must be much larger than the fluid density $\rho_f$,
so that buoyancy and pressure gradient force of the undisturbed flow can be neglected.
Also, brownian motion is assumed to be negligible.
The velocity field $\vec V_f$ is taken to be a steady flow
  submitted to a weak unsteady perturbation~:
$$
\vec V_f(\vec x,t) = \vec V_f^0(\vec x) + \eps\, 
\vec V_f^1(\vec x,t),\quad\quad \eps \ll 1.
$$
Following Fung \cite{Fung1997} we will consider the case
where $\vec V_f^1$ is $T$-periodic with $T=2\pi/\w$.
If $V_0$ denotes a typical fluid velocity and $L_0$ is the length
scale of velocity gradients in the absence of inclusion
the particle motion equation,
non-dimensionalized by ($V_0$,$L_0$), is :
\beq
\tau\, {\vec{\ddot X}_p} = \vec V_T + \vec V_f^0(\vec X_p) + \eps\, 
\vec V_f^1(\vec X_p,t) - \vec{\dot X}_p
\label{eqmvt}
\eeq
without renaming $t$, $\vec X_p$ and $\vec V_f$. The parameter $\tau$
is the non-dimensional response time, which is assumed to be much
smaller than unity :
$$
\tau  =  \frac{2\rho_p}{9\rho_f}\, \frac{\mbox{a}^2}{\nu} \frac{V_0}{L_0}  
\quad  {\ll 1}
$$
and $\vec V_T$ is the  non-dimensional terminal
velocity in a fluid at rest (
$\displaystyle{
\vec V_T  =\frac{2 \rho_p}{9\rho_f}\,  \frac{\mbox{a}^2 \vec g}{\nu}\, \frac{1}{V_0}
}$),
which is assumed to satisfy :  
$$
V_T = |\vec V_T| = O(1).
$$
The  Froude number $\tau/V_T = V_0^2/g L_0$ is therefore much smaller 
than unity. Also, the terminal particle Reynolds number 
is of order
$(\a/L_0)\,V_0 L_0/\nu$, and is therefore very small provided the flow Reynolds
number $V_0 L_0/\nu$ is not too large.
The particle motion equation contains two independent 
small parameters,
namely $\eps$ and $\tau$. In the following we set
  $\tau = k\,\eps$, where $k$ denotes any parameter held fixed as $\eps \to 0$. 
Then classical asymptotic expansions \cite{Maxey1987} of the form  
$
\vec{\dot X}_p = \vec V_T + \vec V_f^0 
 + \eps \vec V_p^{1} +   O(\eps^2)
$
lead to :
$$
\vec{\dot X}_p = \vec V_T + \vec V_f^0(\vec X_p) 
$$
\beq
+ {\eps
\Big( \vec V_f^1(\vec X_p,t) -  k\,\tensna \vec V_f^0 . (\vec V_f^0 +\vec V_T)
\Big)}
+ O(\eps^2).
\label{eqmvt2ddl}
\eeq
The case $k=0$ corresponds to heavy particles without inertia
\cite{Maxey1986}\cite{Maxey1987}\cite{Simon1991} : their
velocity is equal to the fluid local velocity plus $\vec V_T$. In this case the 
particle dynamics is non-dissipative since $div\,\vec V_f = 0$.
In the following we will consider the general case $k \ge 0$.
\vs



\section{Formation  of retention zones in plane flow}
\label{H}

As noticed by many authors \cite{Stommel1949}\cite{Maxey1986}\cite{Simon1991}\cite{Fung1997}, 
when the flow
is two-dimensional
 the dynamical system (\ref{eqmvt2ddl})  is, to leading order, hamiltonian  :
\beq
\vec{\dot X}_p = \left(\ba{r} 
\dot x\\
\dot y\ea\right) = \vec V_T + \vec V_f^0(\vec X_p) = \left(\ba{r} 
\frac{\dr H}{\dr y}\\
-\frac{\dr H}{\dr x}\ea\right)\quad 
\label{Ordre0}
\eeq
\vs
where $H(x,y) = \psi^0(x,y) + x \, V_T$, and $\psi^0$ is the streamfunction 
associated
to $\vec V_f^0$, and $x$ is the horizontal  coordinate and $y$  is the upward
vertical coordinate.
This enables to notice the following  property.
Suppose $\psi^0(x,y)$ is of class ${\cal C}^\infty$ and
has a vertical upward streamline \S
(Fig. \ref{BranchesAsc}(a))
and a point C $\in$ \S where the vertical velocity $v = -\dr{\psi^0}/\dr x$ satisfies :
\beqar
\label{condmax1}
\frac{\dr v}{\dr x}\CC &=& \frac{\dr v}{\dr y}\CC = 0\\
\label{condmax2}
\frac{\dr^2 v}{\dr x^2}\CC \frac{\dr^2 v}{\dr y^2}\CC &- &
\Big(\frac{\dr^2 v}{\dr x\dr y}\CC\Big)^2 > 0 \\
\frac{\dr^2 v}{\dr x^2}\CC  &<& 0.
\label{condmax3}
\eeqar
These
 are the classical conditions for $v(x,y)$ to have a  strict 
local maximum at C. We will denote $x_c$ and $y_c$ the coordinates of $C$,
and $\vec x_c = (x_c,y_c)$ its position vector. Also, we write
$V_c = v(C)$ (the local peak flow velocity) and
assume
$V_c > 0$.
 Then for $V_T < V_c$ and $V_T$ sufficiently close to $V_c$,
the particle trajectory at order $\eps^0$ (Eq. (\ref{Ordre0})) takes
the form of  an elliptic dipole of centre C (Fig. \ref{BranchesAsc}(b)). 
If in addition vorticity is zero along \S, the axes of the ellipse 
are $(C,x)$ and $(C,y)$
and the semi-axes are
\beq
a = \sqrt{\frac{6 (V_c - V_T)}{\psi^0_{,xxx}\CC} }
\quad\quad\mbox{and}\quad\quad
b = \sqrt{\frac{2(V_c - V_T)}{\psi^0_{,xyy}\CC} }
\label{ab}
\eeq
(where the coma indicates spatial differentiation).
\bfi
\cl{\includegraphics[height=9cm]{BranchesAscFIG}}
\vspace*{-1.5cm}
\caption{\sl Sketch of the local streamlines {\bf (a)}
of the unperturbed flow $\vec V_f^0$,
and of the  leading order phase portrait of particle dynamics {\bf (b)}
in this flow,
under the
hypotheses listed in the text.
 }
\label{BranchesAsc}
\efi
\vs
This property 
 can be shown by expanding $H(x,y)$ in the vicinity
of C and making use of the fact that many  derivatives of $\psi^0$
vanish at C. First, 
$u(x_c,y) = 0$ for all
$y \in$ \S, so $\dr^n u/\dr y^n (x_c,y) = 0$ for all integer $n$, so 
$\dr^n \psi^0/\dr y^n \CC$ is zero too for all $n$. Also,  velocity
is divergence-free, so $
\psi^0_{,xy}\CC =
u_{,x}\CC =$ $ u_{,x}\CC + v_{,y}\CC = 0$. Also, $\psi^0_{,xx}\CC =
-v_{,x}\CC = 0$. Therefore, by expanding $H(x,y)$ we are led to :
$$
H(x,y) = \psi^0\CC + x_c\,V_T 
+ (x - x_c) 
$$
$$
\times \Big[V_T - V_c
+\frac{\psi^0_{,xxx}}{6}(x-x_c)^2 
+\frac{\psi^0_{,xxy}}{2}(x-x_c)(y-y_c)
$$
\beq
+\frac{\psi^0_{,xyy}}{2}(y-y_c)^2 \Big]
+ {\cal R}
\label{taylor}
\eeq
where all the derivatives of $\psi^0$ have been taken at C and 
${\cal R}= O(|\vec x-\vec x_c|^4)$.
 Therefore,
the isolines 
$H(x,y) = constant$ take {\sl locally} the form of elementary
 curves. In particular,
if ${\cal R}$ is null or negligible,
the line $H(x,y) = H(C)$, for $x \not = x_c$,
is the quadratic curve :
$$
\alpha (x-x_c)^2 
+2\beta (x-x_c)(y-y_c)
+\gamma  (y-y_c)^2 + \delta = 0
$$
where
$$
\alpha = \frac{\psi^0_{,xxx}\CC}{6}, \quad \beta = \frac{\psi^0_{,xxy}\CC}{4},
\quad \gamma = \frac{\psi^0_{,xyy}\CC}{2}
$$
and $\delta = V_T - V_c$.
A sufficient condition for this curve to be an ellipse is~:
$\delta \not = 0$ 
and $\delta (\alpha+\gamma) < 0$
and $\alpha\gamma > \beta^2$.
Assumption (\ref{condmax3}) implies $\alpha > 0$, and by (\ref{condmax2})
we obtain $\gamma > 0$. Also, $V_T < V_c$ means $\delta < 0$. 
The two first ellipticity conditions are therefore fulfilled. The last one
is a consequence of (\ref{condmax2}). Indeed, Eq. (\ref{condmax2}) can 
be re-written as 
$
(6\alpha)(2\gamma) > (4\beta)^2 . 
$
Therefore we have
$
\alpha\gamma > \frac{4}{3}\beta^2,   
$
hence $
\alpha\gamma > \beta^2   
$.
\vs
If in addition vorticity is zero along \S, we have $\psi^0_{,xx} + \psi^0_{,yy} 
= 0$ so $\psi^0_{,xx}=0$
at {\sl any} point in \S. So  $\psi^0_{,xxy}$ is also zero along \S. 
In particular $\beta = 0$, and the ellipse has horizontal and vertical axes.
In this case the semi-axes of the ellipse  are $a=\sqrt{-\delta/\alpha}$
and $b=\sqrt{-\delta/\gamma}$, and this leads to  (\ref{ab}). 

\vs


The remainder ${\cal R}$ is negligible 
 in the limit where $|\vec x-\vec x_c| \ll 1$. On the ellipse this condition
is fulfilled if $\max(a,b) \ll 1$, that is $\mid V_T-V_c\mid \ll 1$, since both $\alpha$
and $\gamma$ are $O(1)$. Therefore, the elliptical structure is asymptotically
valid if $V_T$ is sufficiently close to $V_c$. 
In the following we will assume that ${\cal R}$ is indeed zero or
negligible, for the sake of simplicity.
An analysis of the effect of
${\cal R}$ is presented in Appendix \ref{App2}.
\vs

Under these conditions
 the particle dynamics in the vicinity
of C  has two stagnation points $A=(x_c,y_c+b)$ and 
$B=(x_c,y_c-b)$ located
along the vertical direction, related by three separatrices (heteroclinic trajectories).
These  cells are an example of Stommel's retention zones, already observed 
with various shapes in
many experimental or theoretical works cited in the introduction \cite{Benard1900}
 \cite{Stommel1949}
 \cite{Simon1991} \cite{Fung1997}
\cite{Cerisier2005} \cite{Tuval2005}. 
To leading order $\eps^0$, particles which are located within the elliptic cell are  trapped
and cannot exit. Also, particles released
outside the cell cannot penetrate into the cell : these particles drop. 
Therefore, if one injects particles at random
in the two-dimensional flow with a uniform distribution, 
the percentage of trapped particles should be approximated by
the volume  fraction $\Phi_v$ of the elliptic cells, that is :
\beq
\Phi_v = N_e \, \pi \,a \,b 
=  \frac{N_e\,\pi\sqrt{12}}{\sqrt{\psi^0_{,xxx}\CC\,\psi^0_{,xyy}\CC}}
(V_c - V_T)
\label{Phiv}
\eeq
where $N_e$ is the number of elliptic cells per unit volume. (Since the problem is
two-dimensional, volumes must be understood as surfaces multiplied by the unit 
length along the $z$ direction.) This formula will be checked below (section \ref{appli}).
\vs

If the $O(\eps)$ perturbation appearing in Eq. (\ref{eqmvt2ddl})
is taken into account, the curved separatrix can be broken : particles
could therefore go in and out in a quite unpredictable manner. 
The calculation
of the parameters leading to separatrix splitting is therefore an interesting
topic for the understanding of particle sedimentation, and is done in the next section. 

\section{Heteroclinic bifurcation and chaotic particle motion}
\label{melnik}

For plane flows the particle motion equation (\ref{eqmvt2ddl}) has the form of a 
hamiltonian system with ``one-and-a-half'' degree of freedom. 
The $O(\eps^0)$ terms contain the fluid velocity as well as the sedimentation 
velocity, the combined effect of which can lead to the formation of the elliptic cells
described above. The separatrices of these cells form
  heteroclinic cycles between the stagnation
points A and B.
The $O(\eps)$ terms manifest the effect of both the flow unsteadiness
and particle inertia : clearly, these physical mechanisms might influence 
the particle dynamics in the vicinity of the elliptic cells, since phase portraits
of the form of Fig. \ref{BranchesAsc}(b) can be structurally unstable.

\bfi
\cl{\includegraphics[height=3.5cm]{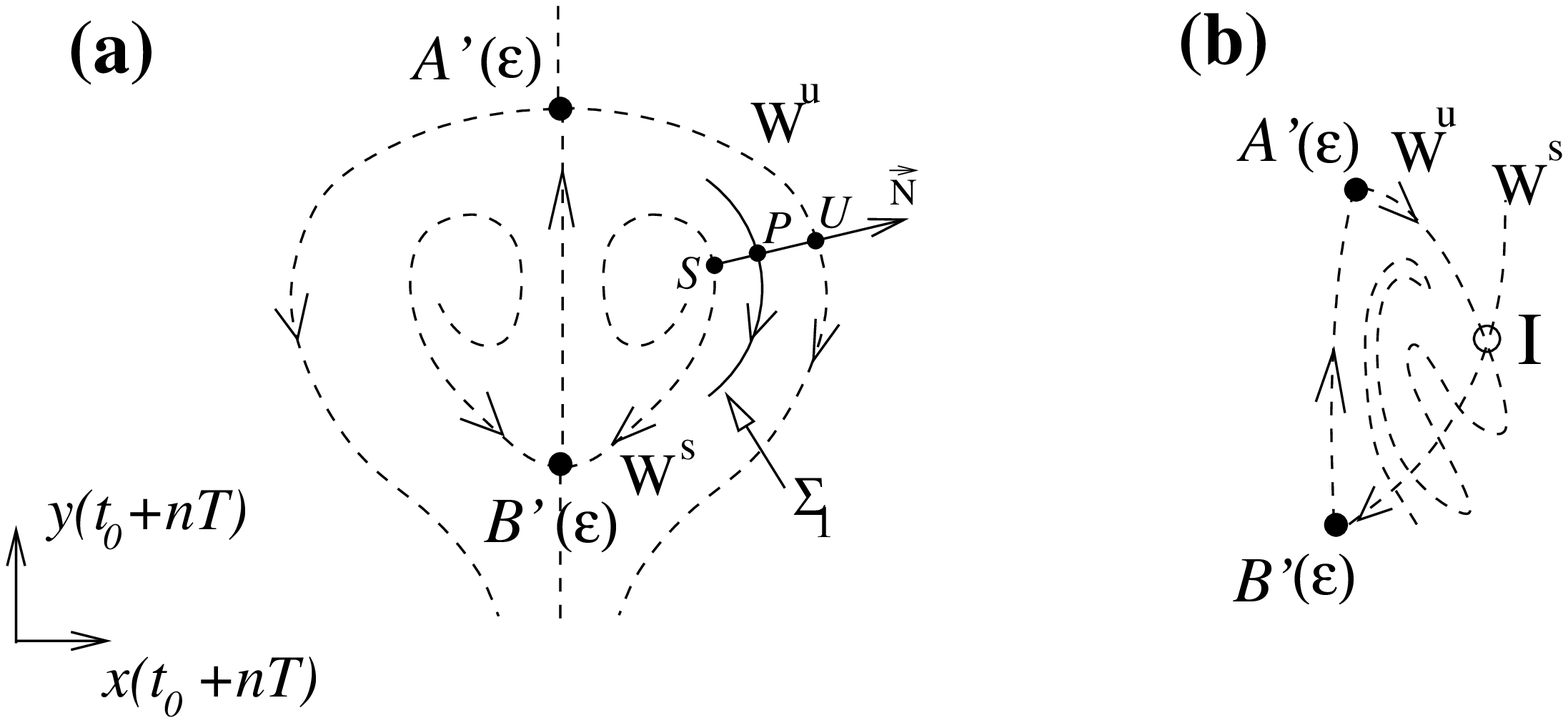}}
\caption{\sl Sketch of the Poincar\'e
section of the
particle dynamics at order  $\eps$. In case (a) the manifolds do not intersect.
Particles released from above the cell will not be trapped. Particles released
within the cell will spiral out and exit the cell.
In case (b) the manifolds intersect and chaotic sedimentation occurs.
 }
\label{BranchesAsc2}
\efi

\vs
In order to investigate analytically the role of the $O(\eps)$ terms,
we make use of the classical Melnikov method \cite{GH01}
to predict the appearance of 
intersection points between the stable and unstable manifolds associated to the
hyperbolic points $A'$ and $B'$ of the {\sl Poincar\'e section} $(x(t_0+nT),y(t_0+nT))$,
 $n \in \Z$,
of the perturbed system
(Fig. \ref{BranchesAsc2}). 
($T$ is the period of the perturbation  velocity field : $\vec V_f^1(\vec x,t+T)
=\vec V_f^1(\vec x)$ for all $\vec x$ and $t$.)
Indeed, since the leading-order particle dynamics
displays two hyperbolic  saddle points $A$ and $B$, the Poincar\'e section
of this system will have hyperbolic points too, at the same position $A$ and $B$.
If $\eps$ is small enough the Poincar\'e section of the {\sl perturbed}
system will therefore have two stagnation  points of the same type 
(hyperbolic-saddle) $A'(\eps)$ and $B'(\eps)$, continuous functions
of $\eps$, by virtue of the implicit functions theorem. Accordingly, there exists
stable and unstable  manifolds W$^s$ and W$^u$ attached to $B'(\eps)$ and $A'(\eps)$
respectively. 
If an unstable manifold intersects a stable one (e.g. point I of Fig.
\ref{BranchesAsc2}(b)), an infinity of such intersection points will exist, since 
 both W$^s$ and W$^u$
are invariant subsets of the Poincar\'e 
application. 
Because the $O(\eps^0)$ phase portrait displays a heteroclinic cycle,
 the particle dynamics in the vicinity of the broken separatrix
 corresponds  to a horse-shoe map,
by virtue of the 
Smale-Birkhoff theorem, characterized by chaotic trajectories and sensitivity to initial 
conditions.
Particles are therefore likely to penetrate
into the cell or exit from it, in a chaotic manner.
\vs
The  occurrence of intersection points can be predicted from the asymptotic calculation
of the ``algebraic distance'' $d(t_0)$
 between  W$^s$ and W$^u$. It is defined as $d(t_0)= \vec{SU}.\vec N$,
where $\vec N$ is the unit vector perpendicular to $\vec V_f^0+\vec V_T$
 at a given (arbitrary) point $P$ of the separatrix,
and such that ($\vec V_f^0+\vec V_T,\vec N,\vec e_z$) is right-handed,
and $S$ (resp. $U$) are the intersection points between the line $(P,\vec N$)
and W$^s$ (resp. W$^u$). 
($\vec e_z$ is the direct unit vector in the direction perpendicular to $(x,y)$.)
These points are shown, for the separatrix $\Sigma_1$,
in Fig. \ref{BranchesAsc2}(a).
At order $\eps$ this
distance is proportional to the Melnikov integral \cite{GH01} :
$$
M(t_0)=\int_{-\infty}^{+\infty} 
\Big[ \vec V_T + \vec V_f^0 \Big]_{\vec X_0(t)} 
\wedge
 \Big[\vec V_f^1\Big(\vec X_0(t),t+{t_0}\Big)
$$
\beq
- k\,
\tensna \vec V_f^0 . \Big(\vec V_f^0 +\vec V_T\Big)_{\vec X_0(t)} \Big]   \, dt
\label{M(t_0)}
\eeq
which is integrated along the separatrix of the unperturbed system, parametrized by
$\vec X_0(t)$ with $\vec X_0(-\infty) = A$ and $\vec X_0(+\infty) = B$ and $\vec X_0(0) = P$. 
(Here, $\vec u \wedge \vec v$ is the triple product $\vec e_z.(\vec u \wedge \vec v)$.)
In particular we have $\vec{\dot X}_0 = \vec V_f^0(\vec{X}_0) +\vec V_T$ and $\vec{\ddot X}_0
=\tensna \vec V_f^0 . (\vec V_f^0 +\vec V_T)$.
 If  particle inertia is taken into account ($k \not = 0$) the perturbation
is dissipative, but this does not alter the validity of Eq. (\ref{M(t_0)}) which only
requires the unperturbed system to be non-dissipative. The Melnikov integral therefore
reads~:
\beq
M(t_0)=\int_{-\infty}^{+\infty} 
\vec{\dot X}_0(t)
\wedge
\vec V_f^1\Big(\vec X_0(t),t+{t_0}\Big) \, dt
 - k\, m
\label{M}
\eeq
where $m = m(V_T)$ is independent of $t_0$ : 
$$
m(V_T) = \int_{-\infty}^{+\infty} \vec{\dot X}_0(t)  \wedge  \vec{\ddot X}_0(t) \, dt.
$$
This term is also independent  of $\vec V_f^1$ and 
corresponds to the effect of
particle inertia. Because $\vec{\dot X}_0  \wedge  \vec{\ddot X}_0$ is 
proportional
to the curvature of the separatrix at $\vec X_0(t)$, the constant $m$
manifests the contribution of centrifugal effects. One can easily check that
$m(V_T) < 0$ (hence $-k m(V_T) > 0$) for the separatrix $\Sigma_1$ of Fig.
\ref{BranchesAsc}(b), and $m(V_T) > 0$ (hence $-k m(V_T) < 0$) for  $\Sigma_3$.
This means that particle inertia tends to maintain $\vec{SU}$ in the direction
of $\vec N$  on separatrix $\Sigma_1$ (that is $\vec{SU}$ points outward),
and in the direction of $-\vec N$
 on separatrix $\Sigma_3$ (that is, $\vec{SU}$ points outward also since $\vec N$
points inward there, as ($\vec V_f^0+\vec V_T,\vec N,\vec e_z$) is right-handed). 
In both cases this means that inertia tends to 
maintain W$^u$ at the periphery of the unperturbed cell and W$^s$ towards the
interior of the cell. 
In particular,
this enables to conclude that in the steady case ($\w=0$) the phase
portrait of the $O(\eps)$ particle dynamics takes the form
of Fig. \ref{BranchesAsc2}(a). The Stommel cell is broken
because of particle inertia only : particles released above the cell will never be
trapped, and particles released inside the cell will always escape :
these results agree with those of  Rubin {\it et al.} \cite{Rubin1995},
which have been obtained for aerosols in a cellular flow field.


In the following, the Melnikov integral is calculated 
in the case where ${\cal R} = 0$
 in Eq. (\ref{taylor}), and 
for the elliptic separatrix $\Sigma_1$ of Fig. \ref{BranchesAsc}(b).
Also, the constant $\beta$ is taken to be zero, for the sake of simplicity.
 The calculation
of $\vec X_0(t)$
 is shown in Appendix \ref{App0} and we
 are led to  :
\beq
m(V_T) = -2\pi\sqrt{3} \sqrt{\frac{\psi^0_{,xyy}\CC}{\psi^0_{,xxx}\CC}} 
\Big(V_T-V_c\Big)^2.
\label{m}
\eeq
We observe that
$m(V_T)$ is proportional to the ratio $a/b$ (if
$V_T-V_c$ is held fixed), so that flat horizontal ellipses ($b \ll a$) will be more 
sensitive to particle inertia than flat vertical ellipses ($b \gg a$), as expected
for a centrifugal effect.

The non-constant part of the Melnikov function
requires to specify the kind of perturbation applied.  In this paper, 
``homothetic'' perturbations of the form (Fung \cite{Fung1997})
\beq
\vec V_f^1(\vec x,t) = \vec V_f^0(\vec x) \, \sin \w t,
\label{selfsimi}
\eeq
will be considered.
This corresponds to flows with variable intensity 
without any change in the shape of the streamlines.
Clearly, there is no chaotic advection of pure tracers in this flow, but it 
will be shown
below that chaotic particle sedimentation will occur.
In the case of a homothetic perturbation  the term 
$\vec V_f^0 \wedge \vec V_f^1$  vanishes, and calculations lead to  
$M(t_0) = -A(\w,V_T) {\cos \w t_0} - k\,m(V_T)$, where $A(\w,V_T)$ is proportional
to the Fourier transform of  the horizontal component
of $\vec X_0(t)$ and reads
(see Appendix \ref{App0})~:
\beq
A(\w,V_T) = \frac{\sqrt{3}}{\sqrt{\psi^0_{,xyy} \, \psi^0_{,xxx}}}
  \frac{\pi\, V_T\, \w} 
{\mbox{cosh}\frac{\pi\w}{2\sqrt{2}\sqrt{\psi^0_{,xyy}(V_c-V_T)}}} ,
\label{Mselfsimi}
\eeq
where derivatives of $\psi^0$ have been taken at C.
Clearly,  if $A(\w,V_T) < k\,\mid m(V_T)\mid$ the Melnikov integral 
does not have any zero, and this provides an  analytical criterion 
 to predict the splitting of the curved
separatrix of the retention zone. 
It also enables to predict the lowest velocity $V_T$ below which no
chaotic trapping can occur (see appendix \ref{App1}). 
Note again that these calculations are expected to be valid either if ${\cal R} \equiv 0$
or if $V_T$ is sufficiently close to $V_c$.
These  results will be used in section \ref{appli} and compared 
to numerical solutions of Eq. (\ref{eqmvt}).

\section{Application to  plane  flows}
\label{appli}

In order to validate the above results we have computed
 numerical particle trajectories for various plane flows,  and checked whether
chaotic settling appears for the parameters predicted by Melnikov's analysis.
The results are particularly accurate when the basic flow is itself a third-order
polynomial function of $(x,y)$, e.g. for a cellular flow of the form
$
\psi^0(x,y) = x (x^2+y^2-1),
$
since the Taylor expansion (\ref{taylor}) is exact there. Results are also 
satisfactory for various non-polynomial flows, provided $\mid V_T-V_c\mid$ is
small enough. Consider for example
 the streamfunction
$
\psi^0(x,y) = \sin x \,\sin y,
$
already investigated by several authors (Maxey \& Corrsin
\cite{Maxey1986},
  Fung \cite{Fung1997}, Rubin {\it et al.} \cite{Rubin1995}).
 One can easily check
that there is an array of upward vertical 
streamlines in the unperturbed velocity field, for example :
 $x=0$ and  $-\pi \le y \le 0$. 
The vorticity is zero on this streamline and
the upward velocity has a maximum at
C = $(0,-\pi/2)$. Also, one can easily check that 
Eqs. (\ref{condmax1})-(\ref{condmax2})-(\ref{condmax3}) are satisfied 
at C = $(0,-\pi/2)$, and that $V_c = 1$.
So  if $V_T < 1$, and $V_T$ close to 1, the particle trajectories at order
$\eps^0$ in the vicinity of C
take the form of elliptic dipolar vortices  with half-axes given by Eq. (\ref{ab})  :
$$
a = \sqrt{6(1-V_T)} \quad \mbox{and} \quad b = \sqrt{2(1-V_T)}.
$$
Figure \ref{TypicTrajLdCAmpli0}(a) shows
typical numerical trajectories in this flow when $V_T=0.8$, in the steady case
and with very small particle inertia ($\tau=0.002$). 
 The  Stommel cell
is clearly visible, in agreement with the numerical computations of 
Refs. \cite{Stommel1949}\cite{Maxey1986}\cite{Fung1997}. The ellipse
of semi-axes $a$ and $b$ is also plotted for comparison.
\bfi
\cl{\includegraphics[height=12cm]{TypicTrajTGFIG}}
\vspace*{-2.5cm}
\caption{\sl Upper graph : typical particle trajectories when $V_T=0.8$, $\w=0$, $\eps=0.001$
 and $k=2$
(thick solid lines, numerical computation).
Thin solid lines are flow streamlines. The dashed line is the ellipse
corresponding to the analytical result (\ref{ab}). Some particles have been released above
the cell, others have been released inside the cell.
Lower graph :
trajectories of two particles released slightly
above the cell when $\eps=0.03$, $\tau=0.06$ and
$\w = 0.5$.  }
\label{TypicTrajLdCAmpli0}
\efi
In order to check the theoretical result
(\ref{Phiv}), which is a direct consequence of the appearance of the elliptic cells, 
we have uniformly released 5000 particles  within the domain $[-\pi,\pi]^2$ and 
solved numerically the full motion equation (\ref{eqmvt}) in the steady case ($\w=0$),
and with a finite but small particle response time $\tau$. 
Then particles above the
horizontal line $y=-\pi$ at time $t$ 
have been called ``suspended''. The percentage of suspended particles 
for large $t$ 
is expected to be close to  the volume fraction of the elliptic cells $\Phi_v$.
Figure \ref{PhivVT0_8_tutu} shows this percentage versus $t$, together with 
$\Phi_v$ (Eq. (\ref{Phiv}) with $N_e = 2/(2\pi)^2$ since there are two elliptic cells
in each periodic box), 
for $V_T = 0.8$, and in the cases  $\tau = 0.001$  and 
$\tau = 0.05$. 
Clearly, the percentage of suspended particles is close to
$\Phi_v$ for long times, but keeps on decaying 
 because of particle inertia effects. The decay is almost undistinguishable for 
$\tau = 0.001$ but is clearly visible for $\tau = 0.05$.
Indeed, since $\tau$ is finite in the numerical solution,
the Poincar\'e section of particle dynamics is of the
form of figure \ref{BranchesAsc2}(a), 
since the Melnikov function $M(t_0)=-k m$ is constant and non-zero in the steady
case,
so that
 the elliptic cells are not completely closed and tend to slowly 
empty under the effect of particle 
inertia alone (see also Rubin {\sl et al.}  \cite{Rubin1995}).
\bfi
\cl{\includegraphics[height=10cm]{PhivVT0_8_tutuFIG}}
\vspace*{-2.5cm}
\caption{\sl Decay of the percentage of suspended particles in
a steady cellular flow (numerical solution of Eq. (\ref{eqmvt})),
with $V_T = 0.8$.
 The solid line is the analytical result
(\ref{Phiv}) corresponding to $\tau = 0$. }
\label{PhivVT0_8_tutu}
\efi

In the unsteady case ($\w  > 0$)
the diagram 
$(\w,V_T)$ indicating the occurrence of separatrix splitting is obtained
by plotting the  curve $A(\w,V_T) - |k\,m(V_T)| = 0$ where 
$A(\w,V_T)$ is the amplitude of the Melnikov function (\ref{Mselfsimi}) and $k\,m$ 
is its constant part. This diagram is shown in Fig. 
\ref{diagstabTGselfsimi} in the case ($k=2$).
If $(\w,V_T)$ is located in the zone $A(\w,V_T) - |k\,m(V_T)| > 0$ 
then the separatrix is broken and particles
can penetrate into the elliptic cell in an unpredictable manner :  ``chaotic 
trapping'' occurs. The lowest terminal velocity below
which no chaotic trapping is expected to occur is given by the analytical expression
obtained in Appendix \ref{App1}, and is approximately $V_T \approx 0.76$.
\bfi
\cl{\includegraphics[height=10cm]{DiagStabTGselfsimi_tutuFIG}}
\vspace*{-2.5cm}
\caption{\sl Plot of the bifurcation diagram given by Melnikov's analysis
for  sedimentation in  the plane cellular flow. }
\label{diagstabTGselfsimi}
\efi

In order to compare these predictions of separatrix splitting with numerical
solutions we have solved Eq. (\ref{eqmvt}) with an
implicit Runge-Kutta algorithm, for 500 particles released
at $t=0$ 
 above the
elliptic cell. 
For $t > 0$ these particles drop and go round the elliptic cell, but some of
them are likely to penetrate inside in case of separatrix splitting,
and this will significantly increase the length of their path. 
The computation of each inclusion is stopped if the particle reaches the ``ground'' $y=-\pi$
or if $t=t_f \gg 1$ (here, $t_f=50$).
We then measure the relative normalized averaged particle path length :
\beq
\theta(\w) = \frac{{\cal L}(\w) - {\cal L}(0)}{{\cal L}(0)},
\label{L}
\eeq
where ${\cal L}(\w)$ is the length of the particle path averaged over the particles.
This quantity is good indicator of the occurrence of separatrix splitting here.
Indeed, if no trapping occurs then 
$\theta(\w) \simeq 0$, because all particle paths are rather similar (they look
like those of  figure
\ref{TypicTrajLdCAmpli0}(a)).
If trapping occurs, some particles will spin inside the cell for a while (like in
 figure \ref{TypicTrajLdCAmpli0}(b)) and this will
increase $\theta(\w)$.
\bfi
\cl{\includegraphics[height=10cm]{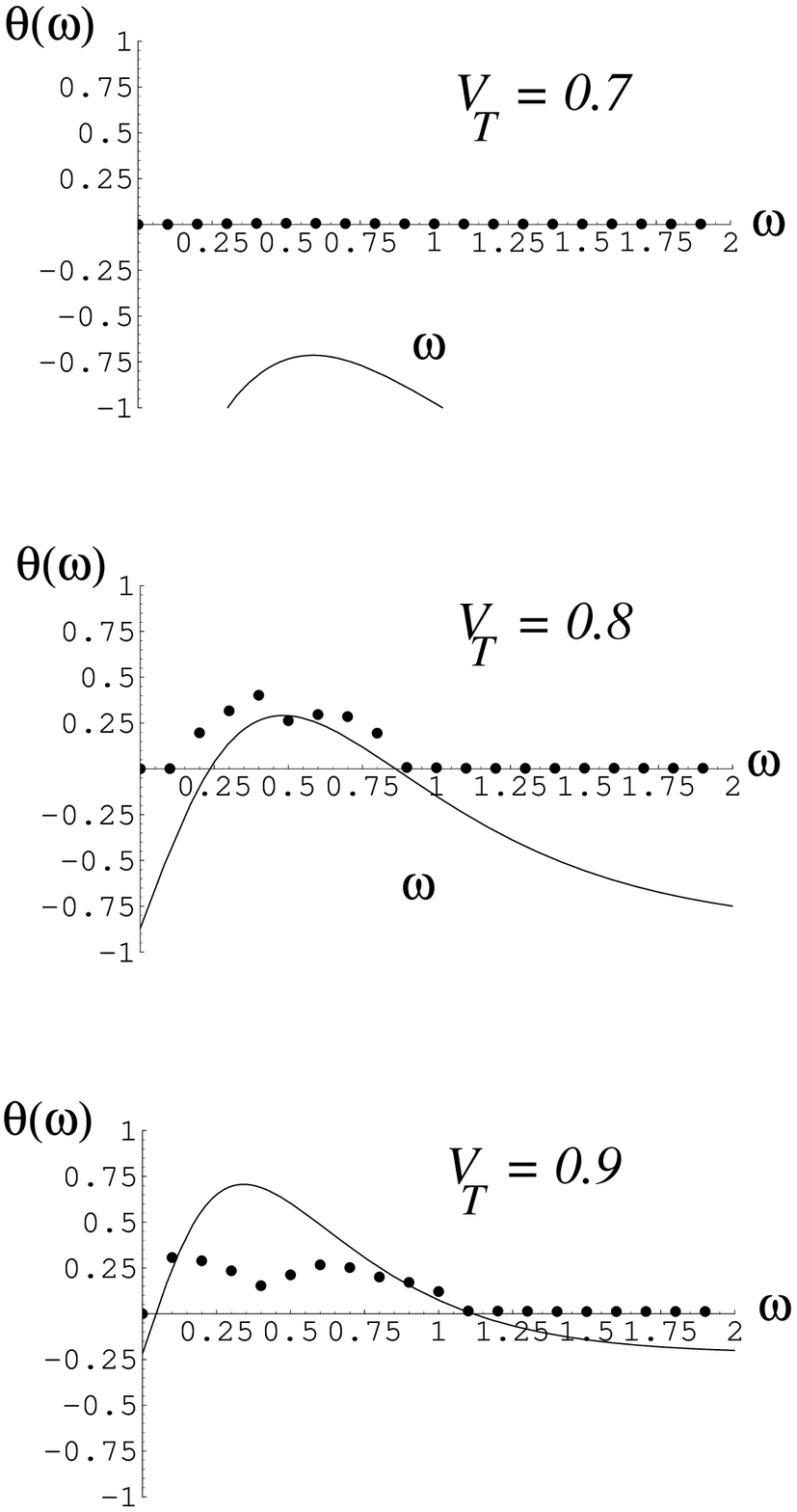}}
\caption{\sl
Plot of the numerical computation of the averaged particle path length in the
unsteady plane cellular
flow $\psi^0=\sin x \sin y$ (black dots),
together with $A(\w,V_T) - k|m(V_T)|$ (solid line,
analytical expressions (\ref{m}) and
(\ref{Mselfsimi})).
The fluid velocity at C is $V_c = 1$, and $\tau = 0.06$, $\eps = 0.03$, ($k=2$).
 }
\label{PourcentTGPlan}
\efi
Fig.
\ref{PourcentTGPlan} shows $\theta(\w)$ versus $\w$ (black dots),
together with $A(\w,V_T)-|k \, m(V_T)|$ (solid line), 
the positive values of which imply separatrix splitting
(here $\eps = 0.03$ and $\tau=0.06$).
Clearly, when $A(\w,V_T)-|k \, m(V_T)| > 0$ some particles are trapped, and when 
$A(\w,V_T)-|k \, m(V_T)| < 0$
no trapping is observed. This confirms that heteroclinic bifurcations do play
an important role in the particle sedimentation process considered here.
\vs
Note that for this flow 
 one can easily check that ${\cal R} = O(\mid\vec x- \vec x_c\mid^5)$
instead of $O(\mid\vec x- \vec x_c\mid^4)$, and it is shown in Appendix (\ref{App2}) that
this allows us to neglect ${\cal R}$ provided
  $\mid V_T - V_c \mid$ is small enough.

\section{Generalization to  axisymmetric flows}
\label{3D}

The analysis presented above can be readily generalized to 
particle sedimentation in an axisymmetric flow displaying an upward streamline \S
along the symmetry axis. If $z$ denotes the coordinate along \S and $r$ is the distance
to \S, the velocity field can be written  :
$$
{\vec V_f^0 . \vec e_r = \frac{1}{r}\frac{\dr \psi^0}{\dr z}}
$$
$$
{\vec V_f^0 . \vec e_z =  -\frac{1}{r}\frac{\dr \psi^0}{\dr r}},
$$
where $\psi^0$ is the streamfunction of the unperturbed velocity field, which is
assumed to satisfy
\beq
\psi^0(r,z) = r^2\, f(r,z)
\label{f}
\eeq
where all the derivatives of $f$ remain finite as $r \to 0$.
The particle velocity in cylindrical coordinates
reads $\vec{\dot X}_p = (\dot r,\dot z) $, and from Eq. (\ref{eqmvt2ddl}) we
obtain the particle dynamics
at order  $\eps^0$  :
$\ds{\dot r = \frac{1}{r}\frac{\dr \psi_p}{\dr z}}$ and 
$\ds{\dot z =  -\frac{1}{r}\frac{\dr \psi_p}{\dr r}}$ ,
where
$$
\psi_p(r,z) = \psi^0(r,z) + {\frac{1}{2} V_T\, r^2}
$$
is the ``particle streamfunction''.
 Suppose there exists a point C ($r=0$,$z=z_c$)
in \S such that
$\vec V_f^0 . \vec e_z$ 
has a strict local maximum. Then, in the vicinity of C, a calculation similar to
the one of section \ref{H} leads to
$$
\psi_p(r,z) \simeq \frac{r^2}{2}\Big(\delta + \alpha r^2 + 2 \beta r (z - z_c) + \gamma (z - z_c)^2        \Big) 
$$
plus higher order terms,
with $\alpha = f_{,rr}(C)$, $\beta = f_{,rz}(C)$,  $\gamma = f_{,zz}(C)$ and $\delta =
V_T-V_c$. Here $V_c = \vec V_f^0 . \vec e_z \CC = -2 f\CC$. 
The conditions ensuring that the the gradient of $\vec V_f^0 . \vec e_z$
vanishes at C and that the hessian  is
definite and negative there, are $f_{,r}\CC = f_{,z}\CC = 0$ and
$$
8\alpha \, \gamma > 9 \beta^2 \quad \mbox{and} \quad \alpha > 0.
$$
Like in section \ref{H}, one can easily check
 that these conditions, together with $V_T < V_c$,
 imply the ellipticity conditions
 $\delta \not = 0$ 
and $\delta  (\alpha+\gamma) < 0$
and $\alpha\gamma > \beta^2$.    The set $\psi_p = 0$ is therefore an ellipsoid,
together with the axis \S.
If, in addition, the curl of $\vec V_f^0$ is equal to
zero along \S, then one obtains
  $\beta = 0$, and the ellipsoid has vertical and horizontal axes. 

\vs

It is convenient to write the particle dynamics, in the vicinity of C, in a hamiltonian
form. This can be readily achieved (in the case $\beta = 0$ for the sake of simplicity) 
by setting $\rho(t) = r^2(t)$, so that
\[
 \left(\ba{r} 
\dot \rho\\
\dot z\ea\right) =  \left(\ba{r} 
\frac{\dr H}{\dr z}\\
-\frac{\dr H}{\dr \rho}\ea\right)\quad 
\]
with
$
H(\rho,z) = \rho \Big( \delta  + \alpha \rho + \gamma  (z - z_c)^2 \Big) + {\cal R}, 
$
where ${\cal R}$ contains higher-order-terms.
One can easily check that the phase portrait $H = constant$,  when 
$\delta < 0$ and
${\cal R} \equiv 0$,  is a
closed cell bounded by a parabolic separatrix $\Sigma_1$  and a straight separatrix
$\Sigma_2$. The Melnikov function of the separatrix $\Sigma_1$, when the flow
is submitted to a  homothetic perturbation (Eq. (\ref{selfsimi})), 
can be calculated by using the same methods as above. After some algebra 
 we obtain 
$M(t_0) = -A(\w,V_T) \cos (\w t_0) - k m(V_T)$, with :
\beq
A(\w,V_T) = \frac{\pi V_T}{f_{,rr}\CC\,f_{,zz}\CC}\frac{\w^2}{\sinh(\frac{\pi \w}{2 \sqrt{f_{,zz}\CC (Vc - V_T)}}) }
\label{A_axi}
\eeq
and 
\beq
m(V_T) = -\frac{8}{3}  \, \frac{\sqrt{f_{,zz}\CC}}{f_{,rr}\CC}\, (V_c - V_T)^{5/2}.
\label{m_axi}
\eeq
(Here also the comma indicates spatial derivation.)
Like for the plane case, the diagram 
$(\w,V_T)$ indicating the occurrence of separatrix splitting is obtained
by plotting the isolines of the curve $A(\w,V_T) - |k\,m(V_T)|$. 
This can be done as soon as the
basic flow $\vec V_f^0$ is given, i.e.
$f_{,rr}\CC$ and $f_{,zz}\CC$ are determined. 
 These results are applied to a pipe flow
in the next section.

\section{Application to particle settling in a vertical straight constricted duct}
\label{appli3D}

Constricted ducts are  investigated in various areas of  fluid mechanics. 
For example, 
the axisymmetric creeping flow through a cylindrical duct with varying cross-section
provides an elementary biomechanical model for stenosed arteria or airways. It is known
that the fluid acceleration through the constricted section can play an important role
in various transfer processes, like particle motion. Here we consider a 
vertical straight constricted duct  (Fig. 
\ref{ArtereStnose}), with a vertical prescribed flow rate $Q$, laden
with heavy microparticles released above the throat of the pipe. 
Under the combined effect of gravity and of the vertical 
flow, particles are likely to drop through the throat, or drop to the throat and get
deviated towards the pipe wall, or get trapped into the throat
if the flow is perturbed. 
\bfi
\cl{\includegraphics[height=6cm]{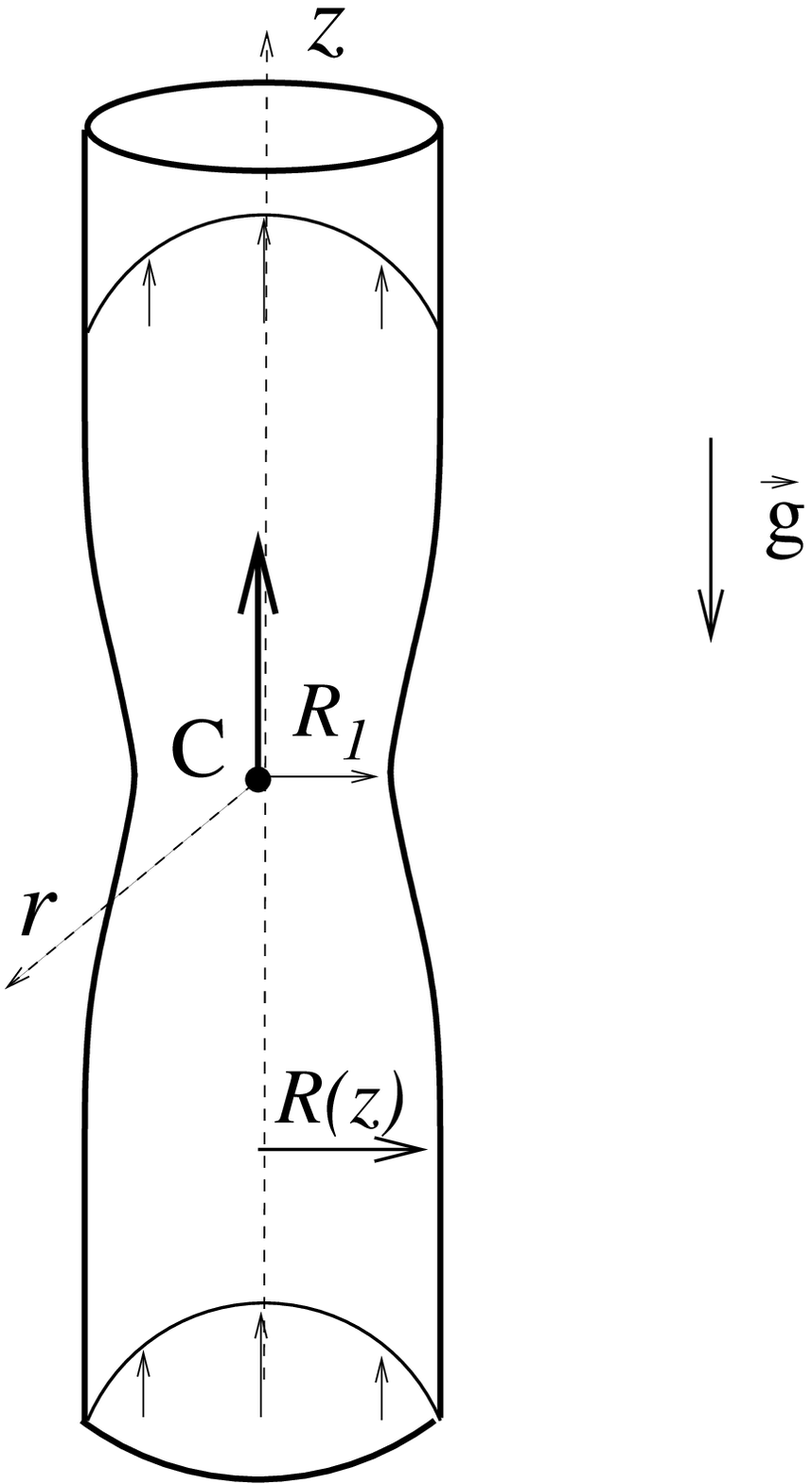}}
\caption{\sl
Sketch of the vertical constricted pipe.}
\label{ArtereStnose}
\efi
The non-dimensional streamfunction is taken to be :
$$
\psi^0(r,z) = -\frac{r^2}{2 R^2(z)} + \frac{r^4}{4 R^4(z)}
$$
where $R(z)$ denotes the pipe radius : $R(0) = R_1 < 1$ at the throat
and $R(\infty) = 1$ far from the throat. It corresponds to a 
parabolic Poiseuille flow in each cross-section of the pipe, with a prescribed
flow rate  $Q = \pi/2$ and a peak axial velocity (at the tip of the parabola)
equal to
 1 far from the throat. This solution is valid if both the pipe Reynolds
number and $1-R_1$ are small.
We have chosen $R(z) = 1-(1-R_1)\exp(-z^2)$, with $R_1 = 0.8$.
At $C = (0,0)$ we have $f_{,r} = f_{,z} = 0$
, $f_{,zz} > 0$, $f_{,rr} > 0$ and  $f_{,zr} = 0$. The conditions required
for $\vec V_f^0 . \vec e_z$ to have a strict local maximum at C are therefore
satisfied, and we conclude that an elliptic Stommel retention zone can form
there for particles such that $V_T < V_c = 1/R_1^2 \simeq 1.56$ 
and $V_T$ close enough to $V_c$.
The complete bifurcation diagram of this flow is
 obtained by  plotting the curve $A(\w,V_T)-|k m(V_T)| = 0$, and
is shown in Fig. 
\ref{DiagStabTuyau} in the case $k=2$. 
\bfi
\cl{\includegraphics[height=6cm]{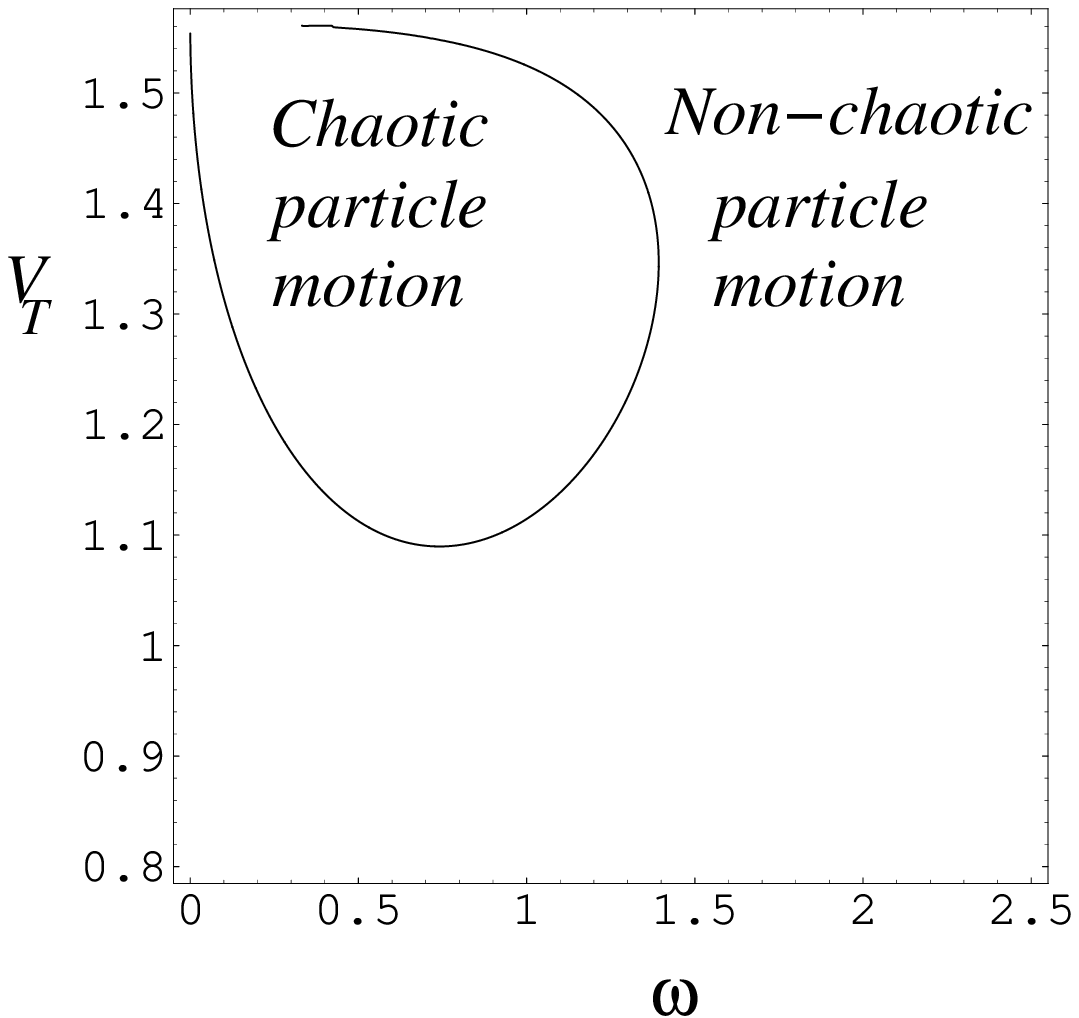}}
\caption{\sl Bifurcation diagram given by Melnikov's analysis
for  sedimentation in the  constricted pipe flow.}
\label{DiagStabTuyau}
\efi

\bfi
\cl{\includegraphics[height=4.7cm]{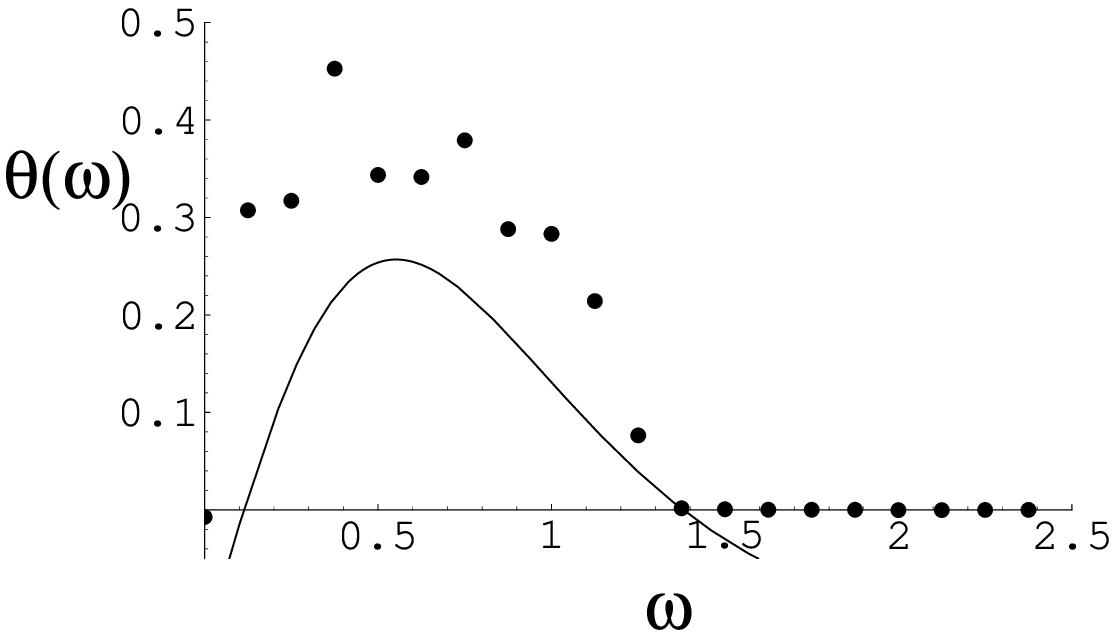}}
\caption{\sl
Plot of the numerical computation of the averaged particle path length in the
unsteady constricted pipe
flow (black dots), together with the analytical expression of
$A(\w,V_T) - k\,|m(V_T)|$ (solid line, Eqs.
(\ref{A_axi}) and (\ref{m_axi})).
The fluid velocity at C is $V_c \simeq 1.56$, and $V_T=1.3$, $\eps = 0.02$, $\tau=0.04$.
The absence of particle trapping  corresponds
to the absence of zeros in the Melnikov function (i.e. $A(\w,V_T) - k\,|m(V_T)| < 0$).
}
\label{w_NpsTuyau}
\efi
These analytical results are compared to
numerical solutions of Eq. (\ref{eqmvt}) in Fig. \ref{w_NpsTuyau},
for $\tau = 2 \eps = 0.04$ and $V_T=1.3$. Like for the plane
case we observe that the absence of particle trapping closely corresponds
to the absence of zeros in the Melnikov function (i.e. $A(\w,V_T) - |k m(V_T)| < 0$). 
This agreement
is observed even though the remainder ${\cal R}$ of the Taylor expansion of $H(\rho,z)$ is not
strictly zero here. 
(Note that, like for the cellular flow investigated above, the remainder
${\cal R}(\rho,z)$ of this model pipe flow is of order 5 instead of  4.) 

In order to visualize the topology of chaotic particle dynamics we 
have plotted a
particle cloud released at $t=0$ at the center of the throat (Fig. \ref{NuagesPubli}),
for the same parameters and with $\w = 0.5$.
Trajectories are obtained by solving numerically Eq. (\ref{eqmvt}).
The stretching and folding of the cloud is clearly visible, 
as expected for chaotic motion.

\bfi
\cl{\includegraphics[height=4.2cm]{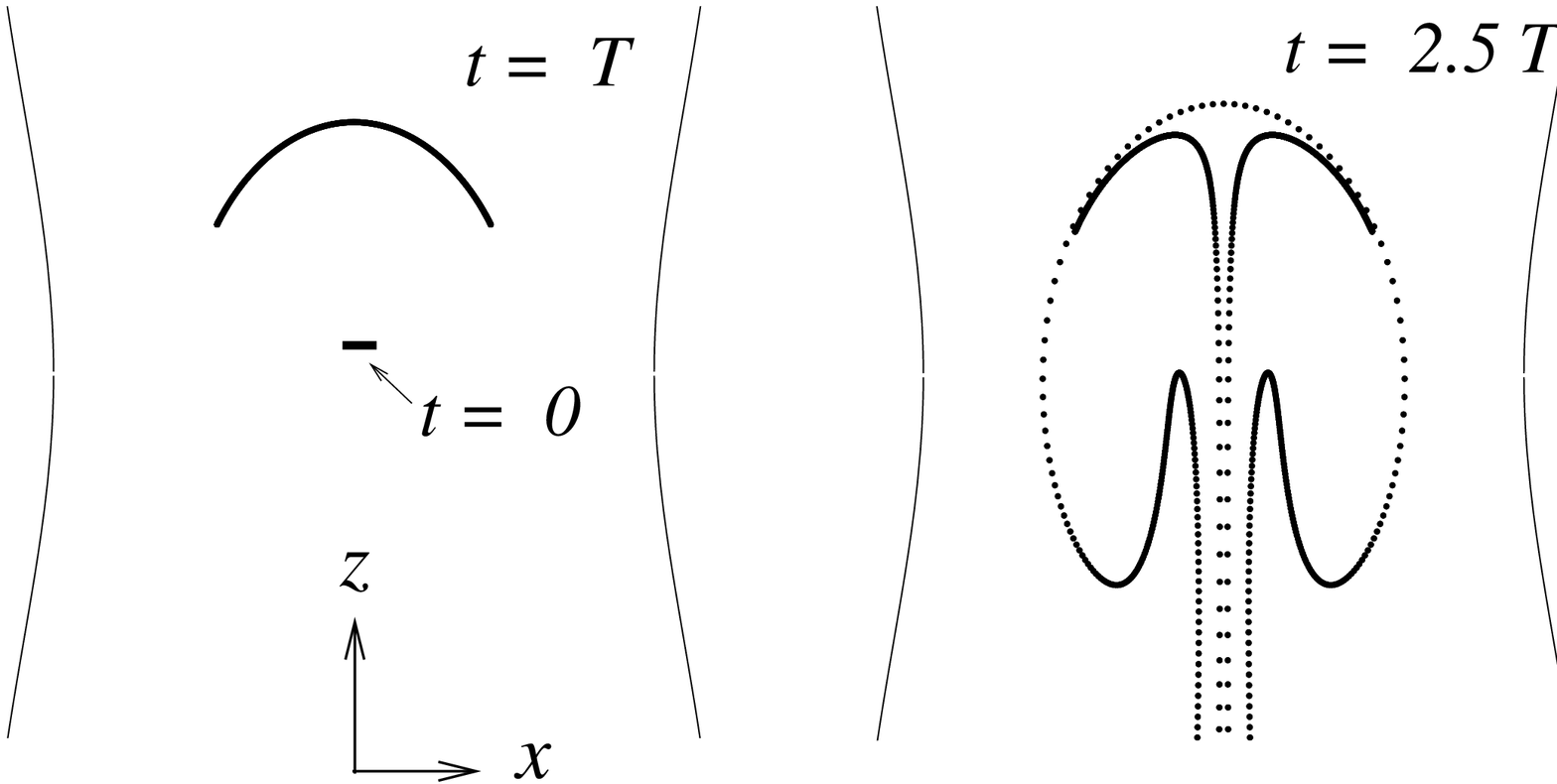}}
\caption{\sl Plot of a particle cloud released at the center of
 the throat,
for the conditions of figure \ref{w_NpsTuyau} and $\w=0.5$.  }
\label{NuagesPubli}
\efi

\section{Conclusion }

The conclusions of this paper are valid  for passive isolated tiny heavy
low-Re particles with a  response time
much smaller than the flow time scales,
 and a terminal velocity of the order of the flow velocity. For such inclusions, sedimentation
effects appear in the leading-order dynamics, whereas particle inertia  can be treated
 as a perturbation, together with  the flow unsteadiness. This enables to write the
particle dynamics as a {\sl non-dissipative} system submitted to a dissipative perturbation.

Elliptic  Stommel cells in steady plane
 or  axisymmetric flows have been shown to occur as soon
as the basic flow has an upward streamline displaying a strict local
maximum with peak velocity $V_c$, and for heavy Stokes particles such that $V_T < V_c$
and $V_T$  close enough to $V_c$, in the limit of vanishing particle inertia. 
 These results are not really new,
as many authors already observed particle trapping in the vicinity of upward streamlines
 \cite{Stommel1949} \cite{Simon1991}
\cite{Cerisier2005} \cite{Tuval2005}. However, the calculations of the present paper confirm that
elliptic Stommel cells can form in a wide variety of flows,
and do not require the flow to have closed streamlines or stagnation points. In particular, they
  can  appear also if the basic flow is an open flow, like the vertical pipe flow with 
varying radius
investigated above.

When the flow is submitted to a weak time-periodic
perturbation, the separatrices of these cells can be broken (heteroclinic bifurcation)
and chaotic trapping, as well as
 chaotic trajectories, can occur. 
The parameters leading to separatrix splitting have been calculated 
in the case where only the quadratic non-linearities of the fluid velocity field can be taken
into account (i.e. ${\cal R}$ is zero or negligible).
We have shown that 
particle inertia  prevents the appearance of chaotic dynamics in the present case,
 by inducing an additive constant $-k\,m$ into the Melnikov function $M(t_0)$.
Therefore, particle inertia tends to maintain the stable and unstable manifolds
W$^s$ and W$^u$ far from each other. This is clearly a centrifugal effect, since the
constant part of the Melnikov function is mainly an integral of the curvature of 
the separatrix. 
In contrast, the flow unsteadiness tends to 
make the manifolds intersect, and the most dangerous perturbation frequency is 
the peak frequency
of the Fourier transform  of a combination of the coordinates of the
point $\vec X_0(t)$ running on the separatrix. This combination depends
on the detailed shape of the perturbation.
The analytical expression of $M(t_0)$ provides efficient criteria to predict 
the occurrence of chaos.
These criteria  have been tested by means of numerical computations, and the agreement is 
satisfactory~: when $\psi^0$ is a cubic function of position (${\cal R}=0$)
 chaotic trapping is observed only if the Melnikov analysis 
predicts separatrix splitting.
The agreement is also satisfactory if ${\cal R}$ is non-zero but decays faster than $\mid
\vec x-\vec x_c\mid^5$ 
provided $\mid V_T-V_c \mid$ is small enough.

\vs

The Melnikov function  $M(t_0)$ of
the {\sl vertical} separatrix $\Sigma_2$ has no constant part : $m(V_T) = 0$ for
all $V_T$. Furthermore, 
for the homothetic perturbation 
(\ref{selfsimi}) we have
 $M(t_0) = 0$ for all $t_0$, so that the first-order Melnikov analysis is 
not conclusive there. In contrast, other kinds of perturbations, 
like periodical rotation around C or translation, lead to  Melnikov functions
of the form $M(t_0) = A(\w,V_T)\sin(\w t_0 + \phi)$ which enable to conclude that $\Sigma_2$
is also splitted : this leads to extremely efficient mixing properties which could be of
 interest, for example,
in microfluidic devices. Indeed, if the three  separatrices break, particles released
in the right-hand-side half-plane ($x > 0$) could penetrate into the cell
and get mixed with particles initially located in the left-hand-side
 half-plane ($x < 0$).

\vs

As noted before, the  elliptic cells, like the  heteroclinic bifurcation,
can occur in a wide variety of flows. We have checked that if the basic flow is 
itself an elliptic dipole (like  Hadamard cells within a bubble) 
with center C, then particle trajectories
can take the form of a  smaller elliptic dipole centred at C. 
When this basic flow is submitted to a
time periodic perturbation, separatrices can be broken and particles can have 
 complex trajectories within the dipole : some remain trapped in the Stommel
cell, and some exit the cell and drop to the bubble interface.
 This model
can be applied to dust transport within a drop or a bubble, a topic
of interest in process engineering, and could help calculate bubble contamination
rates. 
Also, for
 inner Hadamard cells or Hills' vortex 
one can check that 
${\cal R} \equiv 0$, 
so that the Taylor expansion of the Hamiltonian $H$ is exact there, 
and the undergoing analytical results are particularly accurate.

\vs

Finally, for non-heavy particles,
 buoyancy and pressure gradient and  added mass can be readily taken into account
 in the particle motion equation. This leads to interesting particle behaviours which
are currently under study. Also, Basset's drag correction, manifesting the effect
of the unsteadiness of the disturbance flow due to the particle, 
could influence the particle
dynamics and the occurrence of separatrix splitting. One can
check that this correction induces an $O(\eps^{3/2})$
integral term into the asymptotic motion equation (\ref{eqmvt2ddl}) (see also 
Druzhinin \& Ostrovsky \cite{Druzhinin02}). This time integral makes the analysis 
much more complicated, and little results are available there. Note however that
first-order fluid inertia effects, responsible for lift or drag corrections,
 should also be taken into account, since
the Stokes number of the inclusion (measuring the unsteadiness of the particle-induced
flow) is often of the order of  its Taylor number (which characterizes
first-order fluid inertia effects). Unfortunately, no general particle motion
equation is available in this case, and the dynamics of such inclusions remains 
an open issue.

\vskip.3cm
\vs
{\large\bf Acknowledgement}
\vs
The author wishes to thank  J.-P. Brancher  for 
stimulating discussions about fluid dynamics
and dynamical systems.
Remarks from scientists of the Fluid Dynamics group of LEMTA
are gratefully acknowledged.



\appendix

\section{Calculation of the Melnikov function}
\label{App0}

The calculation of the complete integral (\ref{M(t_0)}) requires to solve
the differential equation 
\beq
\vec{\dot X}_0 = \vec V_f^0(\vec X_0(t)) + \vec V_T,
\label{X0}
\eeq 
with
$\vec X_0(-\infty) = A$ and $\vec X_0(+\infty) = B$. Also,
$\vec X_0(t)$ belongs to the {\sl curved} separatrix, so
$$
\alpha (x_0(t)-x_c)^2 
+\gamma  (y_0(t)-y_c)^2 + \delta = 0,
$$
where $\vec X_0 = (x_0,y_0)$.
Therefore, the vertical coordinate of $\vec X_0$ satisfies :
$$
\dot y_0 = -\frac{\dr H}{\dr x} = 2\alpha (x_0(t)-x_c)^2 
= -2\Big( \gamma(y_0(t)-y_c)^2+\delta\Big),
$$
and the solution of this elementary equation, with $y_0(\pm\infty)=y_c\mp b$
and $y_0(0)=y_c$, is 
$
y_0(t) = y_c - b\,\tanh(2\gamma\,b\,t)
$
where $b = \sqrt{-\delta/\gamma}$.
We readily obtain the horizontal coordinate of $\vec X_0(t)$ :
$
x_0(t) = x_c + a/\cosh(2\gamma\,b\,t),
$
where $a = \sqrt{-\delta/\alpha}$.
The constant part of the Melnikov function is therefore
$$
m = \int_{-\infty}^{+\infty} (\dot x_0 \ddot y_0 - \ddot x_0 \dot y_0)\, dt 
$$
$$
= -4 a b^3 \gamma^2 
\int_{-\infty}^{+\infty} \frac{ds}{\cosh^3 s} = -2\pi a b^3 \gamma^2 .
$$
By writing $a$, $b$ and $\gamma$ in terms 
of $V_c$, $V_T$ and $\psi^0$ we are led to expression (\ref{m}).

\vs
The  Melnikov integral corresponding to the homothetic perturbation (\ref{selfsimi}),
by
using (\ref{X0}), reads :
$$
M(t_0)=
\int_{-\infty}^{+\infty} 
\vec{\dot X}_0
\wedge
\vec V_f^0\Big(\vec X_0(t)\Big) \sin \w (t+{t_0}) \, dt
 - k\, m.
$$
Using once again (\ref{X0}) :
$$
M(t_0)=
-\int_{-\infty}^{+\infty} 
\vec{\dot X}_0
\wedge
\vec V_T\, \sin \w (t+{t_0}) \, dt
 - k\, m
$$
$$
= \mbox{Im}\, V_T\,e^{i\w t_0}\, \int_{-\infty}^{+\infty} \dot x_0(t) e^{i\w t} \, dt - k\, m,
$$
where Im denotes the imaginary part and $\vec V_T = - V_T\,\vec e_y$.
By inserting the analytical expression of $x_0(t)$ into this Fourier integral we are led
to :
$$
\int_{-\infty}^{+\infty} \dot x_0(t) e^{i\w t} \, dt = \frac{1}{2 i} \frac{\pi a \w}{b\gamma}
\frac{1}{\cosh(\pi \w/4 b \gamma)},
$$
and this leads to $M(t_0) = -A(\w,V_T) \cos\w t_0 - k\,m(V_T)$, with $A(\w,V_T)$ given by (\ref{Mselfsimi}).

\section{Lower bound for $V_T$ leading to heteroclinic bifurcation}
\label{App1}

In this appendix we make use of the Melnikov function $M(t_0)$ to derive
an analytical expression of the range of terminal velocities $V_T$
for which the heteroclinic bifurcation does not occur.
By using (\ref{Mselfsimi}) we obtain a {\sl necessary} condition for the existence
of both the elliptic structure and simple zeros in $M(t_0)$  :
\beq
V_c  - D(\psi^0) < V_T < V_c,
\label{Vmin}
\eeq
where $D(\psi^0)$ is   :
\beq
D(\psi^0) = \frac{s^2}{3} - \frac{6s^2 V_c-s^4}{3 D_1^{1/3}} + \frac{D_1^{1/3}}{3}
\label{P3_1}
\eeq
and  $D_1 = D_1(\psi^0)$ is given by :
$$
D_1(\psi^0) =
$$
\beq
 27 V_c^2 s^2/2 -9 V_c s^4 + s^6+ \frac{3}{2} \sqrt{81 V_c^4 s^4-12 V_c^3 s^6}
\label{P3_2}
\eeq
and
\beq
s = \frac{\kappa \sqrt 2}{k \pi \sqrt {\psi^0_{,xyy}\CC}},
\label{a_2D}
\eeq
and $\ds{\kappa = 
\max_{q}\frac{q}{\cosh q}\simeq 0.663}$.
To obtain this result we  rewrite (\ref{Mselfsimi}) as 
\beq
A(\w,V_T) = \frac{2\sqrt{6}\, V_T}{\sqrt{ \psi^0_{,xxx}}}
 \sqrt{V_c-V_T}
\,\,
F\Big( \frac{\pi\,\w}{2\sqrt{2}\sqrt{\psi^0_{,xyy}(V_c-V_T)}}   \Big)
\label{A}
\eeq
with 
$F(q) = q/\cosh q$. One can easily check that $F$ is a bounded function with 
$\ds{\kappa = 
\max_{q}\frac{q}{\cosh q}\simeq 0.663}$. Hence, whatever  $\w$ the function $F$ in (\ref{A})
is at most equal to $\kappa$, so that condition
$$
\frac{2\sqrt{6}\, V_T}{\sqrt{\psi^0_{,xxx}}}
 \sqrt{V_c-V_T} \,\,\kappa < k\,\mid m(V_T)\mid \,\, 
$$
implies that $M(t_0)$  does not have any zero.
By taking the square of this inequality, which only involves positive quantities,
 we are led to a condition of the form $P_3(V_T) > 0$
where $P_3(X) = (V_c-X)^3-s^2\,X^2$ is a 3rd order  polynomial, and $s$ is given
by (\ref{a_2D}).
One can easily check that  $P_3$ is a decaying function of $X$, for all $X\ge 0$, and that
$P_3(0) = V_c^3 > 0$ and $P_3(V_c) = -s^2\,V_c^2 <0$ : $P_3$ has a real  root
somewhere between 0 and $V_c$. If $V_T$ is smaller than this root, then $P_3(V_T) > 0$ and $M(t_0)$
does not have any zero. An analytical expression of the root can be found
by solving $P_3$ and we are led to
$
V_T <  V_c  - D(\psi^0) 
$
with $D(\psi^0)$ given by Eqs. (\ref{P3_1})-(\ref{P3_2}). 
This is a sufficient condition for non-chaotic sedimentation. 
The opposite of this condition provides a necessary condition for
the heteroclinic tangle to occur.
\vs
In the three dimensional axisymmetric case the same method can
be applied, starting from the Melnikov function given by (\ref{A_axi})
and (\ref{m_axi}). We are led to the same polynomial $P_3$,
so that formulas
(\ref{P3_1}) - (\ref{P3_2}) can be applied, but with a  different
value for $s$ :
\beq
s = \frac{3 \kappa }{2 \pi k \sqrt {f_{,zz}\CC}},
\label{a_3D1}
\eeq
and $\ds{\kappa = 
\max_{q \ge 0}\frac{q^2}{\sinh q}\simeq 1.1046}$. Finally, note that all these
formulas are valid if ${\cal R} \equiv 0$.

\section{Effect of the remainder of the streamfunction}
\label{App2}

The particle dynamics (\ref{eqmvt2ddl})
 can be written as :
$$
\vec{\dot X}_p = \vec V_T + \vec V_f^0(\vec X_p) 
$$
\beq
+ {\eps
\Big( \vec V_f^1(\vec X_p,t)
 -  k\,\tensna \vec V_f^0 . (\vec V_f^0 +\vec V_T)
\Big)} + O(\eps^2)
\label{eqmvt2ddlA}
\eeq
with
\[
\vec V_T + \vec V_f^0    =  \left(\ba{r} 
\frac{\dr H}{\dr y}\\
-\frac{\dr H}{\dr x}\ea\right).
\]
The Taylor expansion  (\ref{taylor}) of the particle streamfunction is of the form 
$
H(x,y) = H_3(\vec x - \vec x_c) + {\cal R}(\vec x - \vec x_c),
$
where $H_3$ is a third-order polynomial function of $\vec x - \vec x_c$ and
${\cal R}$ is a $q$-th order function of $\vec x - \vec x_c$ with $q \ge 4$.
The inertialess particle velocity is therefore affected by the remainder :
\[
\vec V_T + \vec V_f^0     = \ub{\left(\ba{r} 
\frac{\dr H_3}{\dr y}\\
-\frac{\dr H_3}{\dr x}\ea\right)}_{\vec V_2} + \ub{\left(\ba{r} 
\frac{\dr {\cal R}}{\dr y}\\
-\frac{\dr {\cal R}}{\dr x}\ea\right)}_{\vec W_{q-1}}
\]
where $\vec V_2$ is of order 2 and $\vec 
W_{q-1}$ is of order $q-1$ in $\vec x - \vec x_c$. 
The  particle motion equation (\ref{eqmvt2ddlA}) therefore reads :
$$
\vec{\dot X}_p = \vec V_2 
+\eps\Big(\vec V_f^1(\vec X_p,t)- k 
 \tensna \vec V_2 . \vec V_2 \Big)
+ \Big[\ub{\vec W_{q-1}}_{(i)}
$$
\beq
 -  k \ub{\eps (\tensna \vec V_2 . \vec W_{q-1}
+ \tensna \vec W_{q-1} . \vec V_2 + \tensna \vec W_{q-1}. \vec W_{q-1})
}_{(ii)} 
\Big]
\label{v2}
\eeq
where all the terms within the brackets are induced by
${\cal R}$ and have to be negligible compared to
the other terms.
When $\vec x$ is in the vicinity of the ellipse we have 
$\mid \vec x - \vec x_c \mid = O(\max(a,b)) = O(\mid\delta\mid^{1/2})$,
 since both $\psi^0_{,xxx}(C)$
and $\psi^0_{,xyy}(C)$ in Eq. (\ref{ab}) are $O(1)$. 
Therefore, in the vicinity of the ellipse we have :
$$
 \vec W_{q-1}  = O\Big(\mid\delta\mid^{(q-1)/2} \Big)  \ll 
 \vec V_2  = O\Big(\mid\delta\mid \Big),
$$
$$
 \tensna \vec V_2 . \vec W_{q-1} = O\Big(\mid\delta\mid^{q/2} \Big)
  \ll 
 \tensna \vec V_2 . \vec V_2  = O\Big(\mid\delta\mid^{3/2} \Big)
$$
as soon as $\mid\delta\mid \ll 1$.
Therefore  the particle inertia term 
$\eps \tensna \vec V_2 . \vec V_2$  dominates all  the terms in $(ii)$.
Most importantly, the particle inertia term 
$\eps \tensna \vec V_2 . \vec V_2$ also dominates the 
spurious term $(i)$ provided 
$
k \eps \mid\delta\mid^{3/2}
$
dominates  $\mid\delta\mid^{(q-1)/2}$, i.e. 
$$
k \eps \gg\, \mid\delta\mid^{(q-4)/2}.
$$
This condition is satisfied if
 $q > 4$ and if $\delta$ is sufficiently small.
(This conclusion could have also been obtained by normalizing
$(x,y)$ by $(x/\mid\delta\mid^{1/2},y/\mid\delta\mid^{1/2})$
in the particle motion equation.)
In the flow $\psi^0 = \sin x \sin y$ used here we have $q=5$, 
and this allows us to
 neglect the effect of the bracketted terms of Eq. (\ref{v2}) provided
$\delta$ is small enough. The numerical calculations presented here show
that $\delta$ does not need to be very small.

\newpage



\end{document}